\documentclass[reprint,preprintnumbers,aip,showpacs,superscriptaddress,floatfix]{revtex4-1}

\usepackage{palatino}
\usepackage[latin1]{inputenc}
\usepackage{epsf}
\usepackage{amsmath,amssymb}
\usepackage{latexsym}
\usepackage{calc}
\usepackage{color}
\usepackage{shadow}
\usepackage{epsfig}
\usepackage[pdftex, pdfstartview = {FitH}]{hyperref}
\usepackage{graphicx}
\usepackage{float}

\newcommand{\ben}{\begin{equation}}
\newcommand{\een}{\end{equation}}
\newcommand{\bea}{\begin{eqnarray}}
\newcommand{\eea}{\end{eqnarray}}

\def\bR{{\bf R}}

\def\bA{{\bf A}}

\def\dulR{{\underline{\underline{\bf R}}}}
\def\dulr{{\underline{\underline{\bf r}}}}

\begin{document}
\title{Mixed quantum-classical dynamics from the exact decomposition of electron-nuclear motion}
\author{Ali Abedi}
\author{Federica Agostini}
\affiliation{Max-Planck Institut f\"ur Mikrostrukturphysik, Weinberg 2, D-06120 Halle, Germany}
\affiliation{These authors contributed equally}
\author{E.~K.~U. Gross}
\affiliation{Max-Planck Institut f\"ur Mikrostrukturphysik, Weinberg 2, D-06120 Halle, Germany}
\pacs{31.15.-p, 31.50.-x, 31.15.xg, 31.50.Gh}
\begin{abstract}
We present a novel mixed quantum-classical approach to the coupled electron-nuclear dynamics based on the exact factorization of the electron-nuclear wave function, recently proposed in~[A. Abedi, N. T. Maitra, and E. K. U. Gross, Phys. Rev. Lett. 105, 123002 (2010)]. In this framework, classical nuclear dynamics is derived as the lowest order approximation of the time dependent Schr\"odinger equation that describes the evolution of the nuclei. The effect of the time dependent scalar and vector potentials, representing the exact electronic back-reaction on the nuclear subsystem, is consistently derived within the classical approximation. We examine with an example the performance of the proposed mixed quantum-classical scheme in comparison with exact calculations.
\end{abstract}
\maketitle

\section{Introduction}
Among the ultimate goals of condensed matter physics and theoretical chemistry is the atomistic description of phenomena such as vision~\cite{cerulloN2010, schultenBJ2009,  ishidaJPCB2012}, photo-synthesis~\cite{tapaviczaPCCP2011, flemingN2005}, photo-voltaic processes~\cite{rozziNC2013, silvaNM2013, jailaubekovNM2013}, proton-transfer and hydrogen storage~\cite{sobolewski, varella, hammes-schiffer, marx}. These phenomena involve the coupled dynamics of electrons and nuclei beyond the Born-Oppenheimer (BO), or adiabatic, regime and therefore require the explicit treatment of excited states dynamics. Knowing that the exact solution of the complete dynamical problem is unfeasible for realistic molecular systems, as the numerical cost for solving the time dependent Schr\"odinger equation (TDSE) scales exponentially with the number of degrees of freedom, approximations need to be introduced. Usually, a quantum-classical (QC) description of the full system is adopted, where only a small number of degrees of freedom are treated quantum mechanically, while the remaining degrees of freedom are considered as classical particles. There are two major issues concerning this approximation, namely (i) the separation of the dynamical problem, such that the classical approximation can be performed on only a subset of degrees of freedom, and (ii) the interaction of the two subsystems in the approximate picture. Several attempts~\cite{ehrenfest, mclachlan, shalashilin_JCP2009, TSH_1990, kapral-ciccotti, io, ivano, bonellaJCP2005, marxPRL2002, martinezJPCA2000} to propose a solution to such problems have been investigated over the past 50 years and different approaches to QC non-adiabatic dynamics have been derived. However, a final and general solution to this problem is still lacking.

In this paper, we approach the problem from a new perspective, employing the exact factorisation of the time dependent electron-nuclear wave function~\cite{AMG1,AMG2}. In this framework, coupled evolution equations of the two components of the system are derived without employing any approximation. In particular, the nuclear equation has the form of a Schr\"o\-din\-ger equation in which the coupling to the electronic subsystem is taken into account through time dependent vector and scalar potentials in a formally exact way. These potentials represent what is usually referred to as the electronic back-reaction on the nuclear subsystem. Their presence in the nuclear equation is crucial for determining the force that generates nuclear trajectories within the approximate QC treatment of the full problem. Recently, we investigated~\cite{steps,long_steps} the properties of such potentials and studied the classical nuclear dynamics under the influence of the force extracted from them. Here we present a new mixed QC (MQC) scheme to treat the coupled electron-nuclear dynamics that is systematically derived by taking the classical limit of the nuclear motion in the framework of the exact factorisation. The classical nuclear dynamics within this MQC approach is governed by a force that includes the effect of the time dependent vector and scalar potentials in the classical limit. 

\section{Exact factorisation of the electron-nuclear wave function}
A multicomponent system of interacting electrons and nuclei is non-relativistically described by the Hamiltonian 
\begin{equation}
\hat{H}(\dulr,\dulR)=\hat{T}_n(\dulR)+\hat{H}_{BO}(\dulr,\dulR).
\end{equation}
Here, $\hat T_n(\dulR)$ is the nuclear kinetic energy and 
\begin{equation}
\hat H_{BO}(\dulr,\dulR)=\hat T_e(\dulr)+\hat V_{e,n}(\dulr,\dulR)
\end{equation}
is the BO Hamiltonian, containing the electronic kinetic energy and all interactions. Throughout this paper, the coordinates of the $N_{e}$ electrons 
and $N_n$ nuclei are collectively denoted by $\dulr$, $\dulR$. It has been proved~\cite{AMG1,AMG2} that $\Psi(\dulr,\dulR,t)$, the exact solution of the TDSE with Hamiltonian $\hat{H}$, can be exactly factorised as
\begin{equation}\label{eqn: factorisation}
  \Psi(\dulr,\dulR,t)=\Phi_{\dulR}(\dulr,t)\chi(\dulR,t),
\end{equation}
with $\Phi_{\dulR}(\dulr,t)$ and $\chi(\dulR,t)$ being the electronic and nuclear wave functions, respectively. $\Phi_{\dulR}(\dulr,t)$ depends parametrically on the nuclear configuration and satisfies the partial normalisation condition 
\begin{equation}
\int d\dulr\left|\Phi_{\dulR}(\dulr,t)\right|^2=1, \forall\,\dulR,t.
\end{equation}
This condition makes the product~(\ref{eqn: factorisation}) unique, up to within a (gauge-like) $(\dulR,t)$-dependent phase transformation. The evolution of the electronic and nuclear wave functions is determined by the equations
\begin{align}
\left(\hat{H}_{el}-\epsilon(\dulR,t)\right)\Phi_{\dulR}(\dulr,t)&=i\hbar\partial_t \Phi_{\dulR}(\dulr,t)\label{eqn: el eqn} \\  
\hat H_n(\dulR,t)\chi(\dulR,t)&=i\hbar\partial_t \chi(\dulR,t) \label{eqn: n eqn}
\end{align}
where the electronic Hamiltonian 
\begin{equation}
\label{eq:eHamiltonian}
\hat{H}_{el}=\hat H_{BO}+\hat U_{en}^{coup}[\Phi_\dulR,\chi]
\end{equation}
is defined as the sum of the BO Hamiltonian and the electron-nuclear coupling operator,
\begin{align}
 \hat U_{en}^{coup}&[\Phi_\dulR,\chi]=
 \sum_{\nu=1}^{N_n}\frac{1}{M_\nu} \Big[\frac{(-i\hbar\nabla_\nu-{\bf A}_\nu(\bR,t))^2}{2}+\label{eqn: U} \\
 &\Big(\frac{-i\hbar\nabla_\nu \chi}{\chi}+{\bf A}_\nu(\bR,t)\Big)\left(-i\hbar\nabla_\nu-{\bf A}_\nu(\bR,t)\right)\Big],
 \nonumber 
\end{align}
and the nuclear Hamiltonian is 
\begin{equation}
\label{eq:nHamiltonian}
\hat H_n(\dulR,t) = \sum_{\nu=1}^{N_n}\frac{\hat{\widetilde{\mathbf P}}_{\nu}^2}{2M_\nu} + \epsilon(\dulR,t)
\end{equation}
with nuclear momentum operator $\hat{\widetilde{\mathbf{P}}}_{\nu} =-i\hbar\nabla_\nu+{\bf A}_\nu(\dulR,t)$. 
The electronic and nuclear Hamiltonians in Eqs.~(\ref{eq:eHamiltonian}) and~(\ref{eq:nHamiltonian}) contain a time dependent potential energy surface
(TDPES)
\begin{align}\label{eqn: tdpes}
 \epsilon(\dulR,t) = \left\langle\Phi_{\dulR}(t) \right\vert\hat{H}_{el}- i \hbar\partial_t\left\vert\Phi_{\dulR}(t)\right\rangle_\dulr 
\end{align}
and a time dependent vector potential
\begin{align}
 \label{eqn: A}
 {\bf A}_\nu(\dulR,t)=\left\langle\Phi_{\dulR}(t)\right\vert\left.-i\hbar\nabla_\nu\Phi_\dulR(t)\right\rangle_\dulr,
\end{align}
that together with the electron-nuclear coupling operator~(\ref{eqn: U}), mediate the coupling 
between the electronic and nuclear motion in a formally exact way. Here, $\langle \cdot|\cdot|\cdot\rangle_\dulr$ denotes an inner product over electronic variables. Eqs.~(\ref{eqn: el eqn}) and~(\ref{eqn: n eqn}), along with the definitions given in Eqs.~(\ref{eqn: U})~-~(\ref{eqn: A}), present an exact separation of the electronic and nuclear dynamics which maintains the full correlation between the two subsystems as in the TDSE of the complete system. Hence they provide a rigorous starting point for developing practical schemes by introducing systematic approximations. In particular, the nuclear equation~(\ref{eqn: n eqn}) has the appealing form of a Schr\"odinger equation that contains a time dependent scalar potential~(\ref{eqn: tdpes}) and a time dependent vector potential~(\ref{eqn: A}) that uniquely~\cite{Ghosh-Dhara,RGT} (up to within a gauge transformation) govern the nuclear dynamics and yield the nuclear wave function $\chi(\dulR,t)$. Henceforth, the phase freedom will be fixed by adding the additional constraint $\langle\Phi_{\dulR}(t)\vert\partial_t\Phi_{\dulR}(t)\rangle_\dulr=0$.

\section{Quantum-classical equations of motion}
Toward developing a MQC scheme, we first derive classical nuclear dynamics as the lowest $\hbar$-order 
of the nuclear TDSE in Eq.~(\ref{eqn: n eqn}). The wave function $\chi(\dulR,t)$ is written~\cite{van-vleck} as
\begin{equation}
\chi(\dulR,t) = \exp[i\mathcal S(\dulR,t)/\hbar],
\end{equation}
assuming that the complex function $\mathcal S(\dulR,t)$ can be expanded as an asymptotic series in powers of $\hbar$, i.e. $\mathcal S(\dulR,t) = \sum_{\alpha} \hbar^\alpha S_\alpha(\dulR,t)$. When this expression up to within $\mathcal O(\hbar^0)$ terms is inserted in Eq.~(\ref{eqn: n eqn}), the Hamilton-Jacobi equation~\cite{goldstein} is recovered
\begin{equation}\label{eqn: HJE}
 -\partial_t S_0(\dulR,t) = H_n\left(\dulR,\left\lbrace\nabla_\nu S_0(\dulR,t)\right\rbrace_{\nu=1,N_n},t\right),
\end{equation}
if we identify $S_0(\dulR,t)$ with the classical action and, consequently, $\nabla_\nu S_0(\dulR,t)$ with the $\nu$th nuclear momentum. The classical Hamiltonian in Eq.~(\ref{eqn: HJE}) is
\begin{equation}\label{eqn: classical hamiltonian}
 H_n = \sum_{\nu=1}^{N_n}\frac{\left[\nabla_\nu S_0(\dulR,t)+{\bf A}_\nu(\dulR,t)\right]^2}{2M_\nu}+ \epsilon(\dulR,t).
\end{equation}
The canonical momentum, analogous to the case of a classical charge moving in an electromagnetic field, is
\begin{equation}\label{eqn: canonical momentum}
 \widetilde{\bf P}_\nu(\dulR,t) = \nabla_\nu S_0(\dulR,t)+{\bf A}_\nu(\dulR,t)
\end{equation}
and the classical trajectory is determined by Newton's equation~\cite{chin}
\begin{align}
 \dot{\widetilde{\bf P}}_\nu =- \nabla_{\nu}\epsilon+ \partial_t{\bf A}_\nu-{\bf V}_{\nu}\times{\bf B}_{\nu\nu}+\sum_{\nu'\neq\nu} {\bf F}_{\nu\nu'},\label{eqn: classical evolution}
\end{align}
with ${\bf V}_{\nu}=\widetilde{\bf P}_\nu/M_\nu$. Eq.~(\ref{eqn: classical evolution}) is derived by acting with the gradient operator $\nabla_{\nu}$ on Eq.~(\ref{eqn: HJE}) and by identifying the total time derivative operator as $\partial_t+\sum_{\nu'}{\bf V_{\nu'}\cdot \nabla_{\nu'}}$. Henceforth, all quantities depending on $\dulR,t$ become functions of $\dulR^{c}(t)$, the classical path along which the action $S_0(\dulR^{c}(t))$ is stationary. The first three terms on the RHS of Eq.~(\ref{eqn: classical evolution}) produce the electromagnetic force due to the presence of the vector and scalar potentials, with ``generalised'' magnetic field
\begin{equation}\label{eqn: generalised magnetic field}
{\bf B}_{\nu\nu'}\left(\dulR^{c}(t)\right)=\nabla_{\nu}\times {\bf A}_{\nu'}\left(\dulR^{c}(t)\right).
\end{equation}
The remaining term
\begin{align}
 {\bf F}_{\nu\nu'}&\left(\dulR^{c}(t)\right)= -{\bf V}_{\nu'}\times{\bf B}_{\nu\nu'}\left(\dulR^{c}(t)\right) \label{eqn: off-diagonal force}\\
  +&\left[\left({\bf V}_{\nu'}\cdot\nabla_{\nu'}\right){\bf A}_{\nu}\left(\dulR^{c}(t)\right)- \left({\bf V}_{\nu'}\cdot\nabla_{\nu}\right){\bf A}_{\nu'}\left(\dulR^{c}(t)\right)\right]\nonumber
\end{align}
is an inter-nuclear force term, arising from the coupling with the electronic system. Eq.~(\ref{eqn: off-diagonal force}) shows the non-trivial effect of the vector potential on the classical nuclei~\cite{lu1,lu2}, as it not only appears in the bare electromagnetic force, but also ``dresses'' the nuclear interactions. In cases where the vector potential is curl-free, the gauge can be chosen by setting the vector potential to zero, then Eqs.~(\ref{eqn: generalised magnetic field}) and~(\ref{eqn: off-diagonal force}) are identically zero. Only the component of the vector potential that is not curl-free cannot be gauged away. Whether and under which conditions $\mbox{curl}\,\bA_\nu(\dulR,t)=0$ is, at the moment, the subject of investigations.~\cite{CI_MAG}

The nuclear wave function appears explicitly in the definition of the electron-nuclear coupling operator~(\ref{eqn: U}). Therefore, according to the previous discussion, the approximation
\begin{equation}\label{eqn: nabla chi /chi}
 \frac{-i\hbar\nabla_\nu\chi(\dulR,t)}{\chi(\dulR,t)} =\nabla_\nu S_0\left(\dulR^{c}(t)\right)+\mathcal O(\hbar)
\end{equation}
will be adopted. It will appear clear later that such term in the electronic equation is responsible for the non-adiabatic transitions induced by the coupling to the nuclear motion, as other MQC techniques, like the Ehrenfest method or the trajectory surface hopping,~\cite{tully,barbatti,drukker} also suggested. Here we show that this term can be derived from exact equations, but it represents only the zero order contribution in a $\hbar$-expansion. Moreover, this coupling expressed via $\nabla_\nu S_0(\dulR^{c}(t))$ is \textit{not} the canonical momentum appearing in the classical Hamiltonian (whose expression is given in Eq.~(\ref{eqn: canonical momentum})).

We now introduce the adiabatic basis $\lbrace\varphi_\dulR^{(j)}(\dulr)\rbrace$, the set of eigenstates of the BO Hamiltonian with eigenvalues $\epsilon_{BO}^{(j)}(\dulR)$, and we expand the electronic wave function on this basis
\begin{equation}
\Phi_\dulR(\dulr,t) =\sum_jC_j(\dulR,t) \varphi_\dulR^{(j)}(\dulr).
\end{equation}
The electronic equation~(\ref{eqn: el eqn}) gives rise to an infinite set of coupled partial differential equations for $C_j(\dulR,t)$, containing all coefficients and their first and second spatial derivatives. However, the spatial dependence of the coefficients is negligible when the nuclear wave packet becomes infinitely localised at the classical positions (the density of a classical point particle is a $\delta$-function centred, at each time, at the classical position evolving along the trajectory). Indeed, when the classical approximation strictly applies, the delocalisation or the splitting of a nuclear wave packet is negligible. Therefore, any $\dulR$-dependence can be ignored and only the instantaneous classical position becomes relevant. This is the assumption considered here. As consequence of this hypothesis, the coupled equations for the coefficients simplify to a set of ordinary differential equations in the time variable only
\begin{equation}\label{eqn: el ode}
\dot C_j(t)=-\frac{i}{\hbar}[\epsilon^{(j)}_{BO}-\epsilon]C_j(t)+\sum_k C_k(t) U_{jk},
\end{equation}
where all quantities depending on $\dulR$, as $\epsilon^{(j)}_{BO}$, $\epsilon$ and $U_{jk}$, have to be evaluated at the instantaneous nuclear position. The symbol $U_{jk}$ is used to indicate the matrix elements (times $-i/\hbar$) of the operator $\hat U_{en}^{coup}[\Phi_\dulR,\chi]$ on the adiabatic basis. Its expression, introducing the first- and second-order non-adiabatic couplings, ${\bf d}_{jk,\nu}^{(1)}(\dulR)=\langle\varphi_\dulR^{(j)}|\nabla_{\nu}\varphi_\dulR^{(k)}\rangle_\dulr$ and $d_{jk,\nu}^{(2)}(\dulR)=\langle\nabla_{\nu}\varphi_\dulR^{(j)}|\nabla_{\nu}\varphi_\dulR^{(k)}\rangle_\dulr$, is
\begin{align}
U_{jk}&=\sum_\nu\frac{\delta_{jk}}{M_\nu}\left[\frac{i}{\hbar}\left(\frac{{\bf A}_{\nu}^2}{2}+{\bf A}_\nu\cdot \nabla_\nu S_0\right)+\frac{\nabla_{\nu}\cdot {\bf A}_\nu}{2}\right] \nonumber \\
&-\sum_\nu\frac{1}{M_\nu}\left[{\bf d}_{jk,\nu}^{(1)}\cdot\nabla_\nu S_0-\frac{i\hbar}{2}\left(\nabla_{\nu}\cdot{\bf d}_{jk,\nu}^{(1)}-d_{jk,\nu}^{(2)}\right)\right].\label{eqn: U on BO states}
\end{align}
Similarly, the TDPES and the vector potential can be expressed in the adiabatic basis, as
\begin{align}
&\epsilon\left(\dulR^{c}(t)\right)=\sum_j\left|C_j(t)\right|^2\epsilon_{BO}^{(j)}+i\hbar\sum_{j,k}C_j^*(t)C_k(t)U_{jk}
\label{eqn: tdpes on BO states}\\
&{\bf A}_\nu\left(\dulR^{c}(t)\right)=-i\hbar\sum_{j,k}C_j^*(t)C_k(t){\bf d}_{jk,\nu}^{(1)}.\label{eqn: A on BO states}
\end{align}
The electronic evolution equation~(\ref{eqn: el ode}) contains three different contributions: (i) a diagonal oscillatory term, given by the expression in square brackets in Eq.~(\ref{eqn: el ode}) plus the term in parenthesis in the first line of Eq.~(\ref{eqn: U on BO states}); (ii) a diagonal sink/source term, arising from the divergence of the vector potential in Eq.~(\ref{eqn: U on BO states}), that may cause exchange of populations between the adiabatic states even if off-diagonal couplings are neglected; (iii) a non-diagonal term inducing transitions between BO states, that contains a dynamical term proportional to nuclear momentum (first term in the second line of Eq.~(\ref{eqn: U on BO states})), as suggested in other QC approaches,~\cite{tully,barbatti,drukker} and a term containing the second order non-adiabatic couplings. In particular, the dynamical non-adiabatic contribution follows from the classical approximation in Eq.~(\ref{eqn: nabla chi /chi}) and drives the electronic population exchange induced by the motion of the nuclei.

Eqs.~(\ref{eqn: classical evolution}) and~(\ref{eqn: el ode}) suggest a new MQC scheme, beyond Ehrenfest dynamics. The electronic equation~(\ref{eqn: el ode}), which is shown to be norm-conserving by explicit calculation of the time derivative of $\sum_j \vert C_j \vert^2$, contains the TDPES, time dependent vector potential and the electron-nuclear coupling operator that are derived from the exact equation~(\ref{eqn: el eqn}) and determines the evolution of the electronic subsystem. Hence, Eqs.~(\ref{eqn: U on BO states})~-~(\ref{eqn: A on BO states}) properly account for the coupling between the quantum (electrons) and the classical (nuclei) subsystems. The classical Hamiltonian~(\ref{eqn: classical hamiltonian}) governs the dynamics of the nuclear subsystem and contains the scalar and vector potentials representing the quantum back-reaction of electronic non-adiabatic transitions on nuclear motion.

\section{Non-adiabatic charge transfer}
We employ this new MQC scheme to study a simple model for which the exact numerical solution is achievable. The original model was developed by Shin and Metiu~\cite{metiu} to study a the non-adiabatic charge transfer processes and consists of three ions and a single electron. 
\begin{figure}[h!]
  \includegraphics*[width=.4\textwidth]{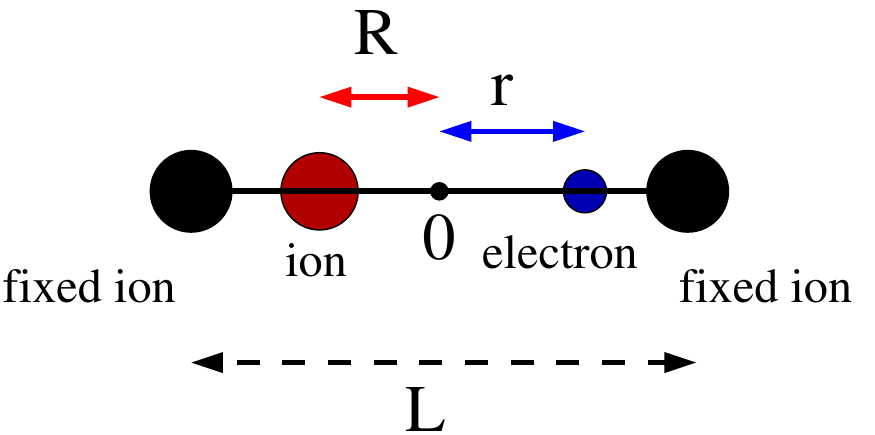}
  \caption{Model system described by the Hamiltonian~(\ref{eqn: metiu}).}
  \label{fig: Figure0} 
 \end{figure}
Two ions are fixed at a distance $L=19.0$~a$_0$, the third ion and the electron are free to move in one dimension along the line joining the two fixed ions. A schematic representation of the system is shown in Fig.~\ref{fig: Figure0}. The Hamiltonian of  this system reads
\begin{align}
 \hat{H}(r,R) =  - \frac{1}{2}\frac{\partial^2}{\partial r^2}-\frac{1}{2M}\frac{\partial^2}{\partial R^2}  + \frac{1}{\vert \frac{L}{2} -R \vert } + \frac{1}{\vert \frac{L}{2} + R \vert} \nonumber\\
 - \frac{\mathrm{erf}\left(\frac{\vert R-r \vert}{R_f}\right)}{\vert R-r \vert} - \frac{\mathrm{erf}\left(\frac{\vert r-\frac{L}{2}
 \vert}{R_r}\right)}{\vert r-\frac{L}{2}  \vert} -
 \frac{\mathrm{erf}\left(\frac{\vert r+\frac{L}{2} \vert}{R_l}\right)}{\vert r+\frac{L}{2} \vert},\label{eqn: metiu}
\end{align}
where the symbols $r,R$ have been used for the positions of the electron and the ion in one dimension. 
Here, $M=1836$, the proton mass, and $R_f=5.0$~a$_0$, $R_l=3.1$~a$_0$ and $R_r=4.0$~a$_0$, such that the first adiabatic potential energy surface, $\epsilon_{BO}^{(1)}$, is coupled to the second, $\epsilon_{BO}^{(2)}$, and the two are decoupled from the rest of the surfaces, i.e. the dynamics of the system can be described by considering only two adiabatic states. The BO surfaces are shown in Fig.~\ref{fig: Figure1a}.
\begin{figure}[!h]
 \includegraphics*[angle=270,width=.3\textwidth]{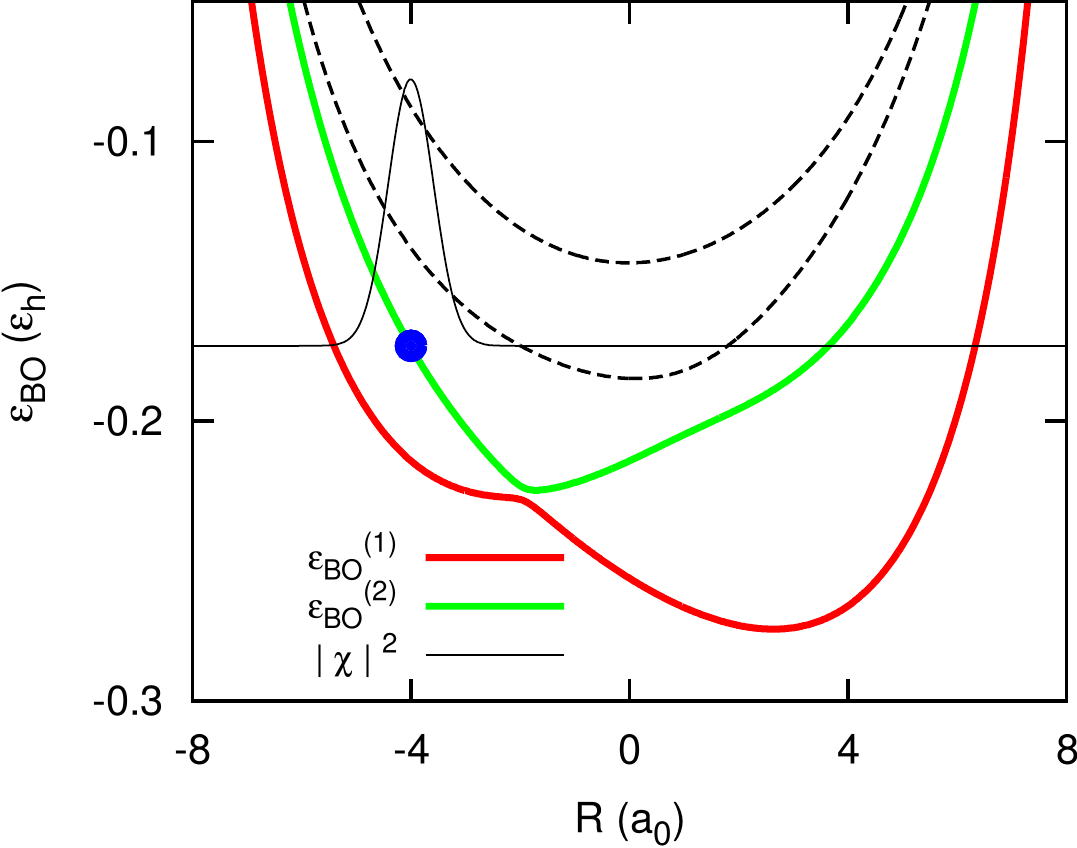}
 \caption{First (red line) and second (green line) BO surfaces, initial Gaussian wave packet (thin black line) centred at $R_0$, indicated as a blue dot (it also indicates the classical initial position). The third and fourth BO surfaces (dashed black lines) are shown for reference.}
 \label{fig: Figure1a} 
\end{figure}

For this model we examine the performance of the MQC scheme in comparison with the exact solution of the TDSE, by using a single-trajectory (ST) and a multiple-trajectory (MT) approaches, referred to as ST-MQC and MT-MQC, respectively. The initial wave function is $\Psi(r,R,0) = G_{\sigma}(R-R_0)\varphi_R^{(2)}(r)$, where $G_{\sigma}$ is a real normalised Gaussian centred at $R_0=-4.0$~a$_0$ with $\sigma=1/\sqrt{2.85}$~a$_0$ and $\varphi_R^{(2)}(r)$ is the second BO state. The classical trajectory starts at $R_0$ with zero initial momentum and $|C_1(0)|^2=0,\,|C_2(0)|^2=1$. If multiple independent trajectories (6000 in this case) are used, initial conditions are sampled according to the Wigner distribution associated to $\Psi(r,R,0)$. We propagate the TDSE numerically with time-step $2.4\times10^{-3}$~fs ($0.1$~a.u.), using the second-order split-operator technique,~\cite{spo} to obtain the full molecular wave function $\Psi(r,R,t)$. The electronic and nuclear equations, in the MQC scheme, are integrated with the same time-step as in the quantum propagation by using the fourth-order Runge-Kutta and the velocity-Verlet algorithm, respectively.

\begin{figure}[!h]
 \includegraphics*[angle=270,width=.3\textwidth]{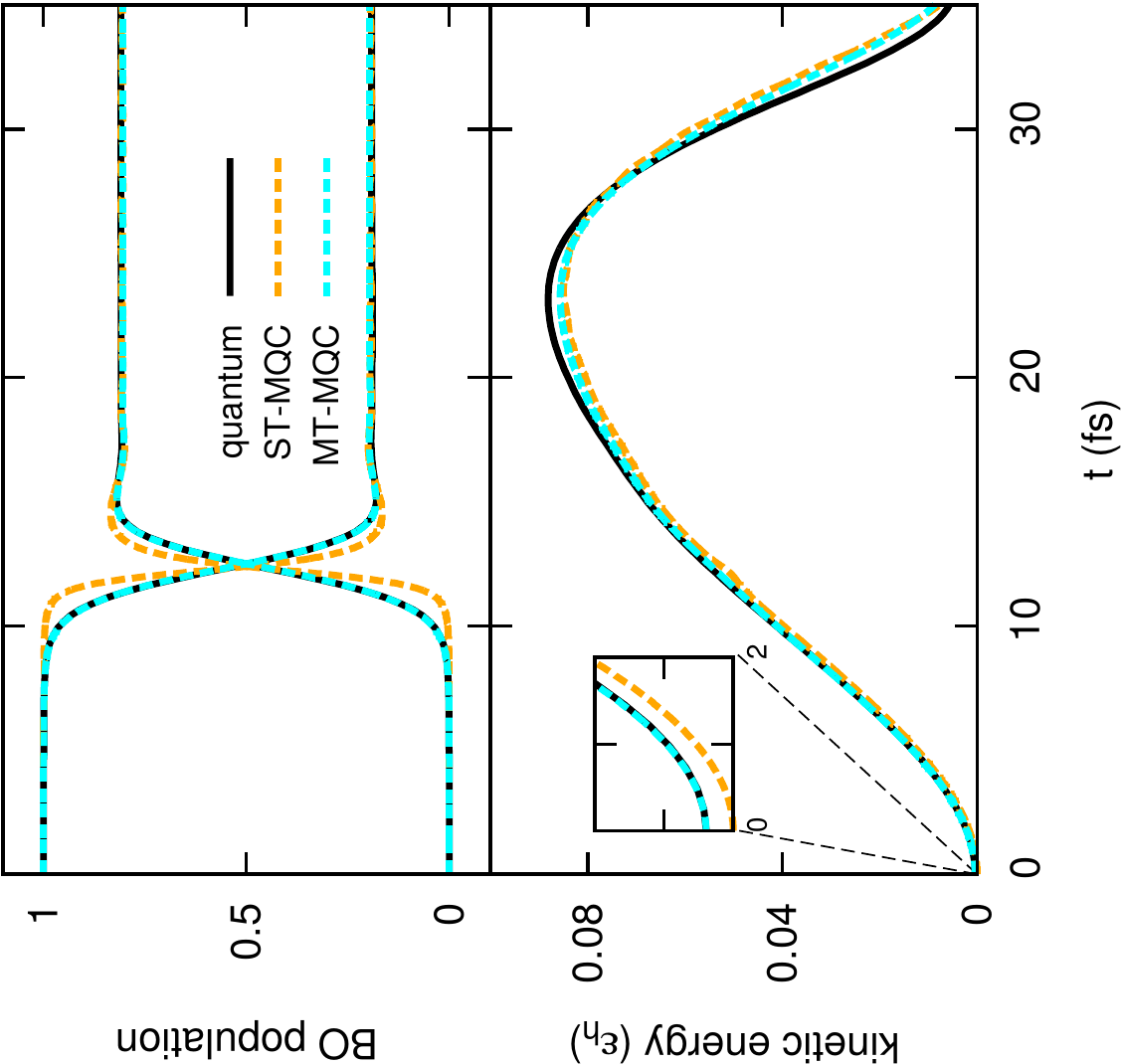}
 \caption{Upper panel: populations of the BO states as functions of time, for exact calculations (continuous black line), ST-MQC (dashed orange line) and MT-MQC (dashed cyan line). Lower panel: nuclear kinetic energy (in Hartree) as a function of time (the color code is the same as in the upper panel).}
 \label{fig: Figure1b} 
\end{figure}
\begin{figure}
\includegraphics*[angle=270,width=.45\textwidth]{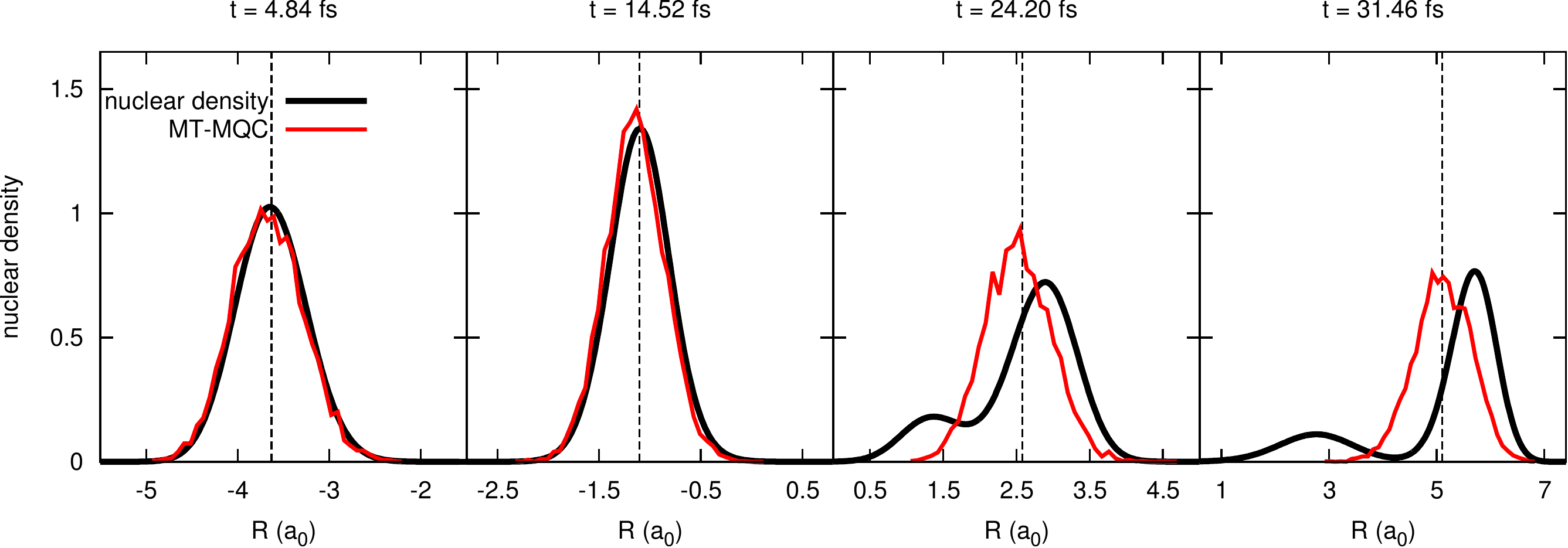}
\caption{Nuclear densities  from exact calculations (black lines) and the MT-MQC scheme proposed here (red lines), at different times as indicated in the plots. The dashed black vertical lines indicate the mean nuclear position from $\Psi(r,R,t)$.}
\label{fig: Figure2}
\end{figure}

The populations of the BO states and the nuclear kinetic energy, as functions of time, calculated from the full electron-nuclear wave function and from the MQC scheme are presented in Fig.~\ref{fig: Figure1b}. It is shown (upper panel) that the MQC evolution (orange line, ST-MQC, and cyan line, MT-MQC) is able to reproduce the branching of the populations of the electronic states after transitioning the avoided crossing at $t\sim 12$~fs, in perfect agreement with the quantum calculations (black line). The use of multiple trajectories allows to smoothen the transition, improving the agreement between 10 and 15~fs. The nuclear kinetic energy (lower panel) from MQC calculations shows a good agreement with exact results, though presenting a slight deviation after the passage through the avoided crossing. It is worth noting that a better agreement with exact calculations is achieved within the MT-MQC scheme at initial (inset in Fig.~\ref{fig: Figure1b}) and final times, where the nuclear kinetic energy is not zero, due to the contribution of the spreading of the quantum nuclear wave packets. The reason of the deviation in the kinetic energy is the spatial splitting of the nuclear density after passing through the avoided crossing, that is not captured by the proposed MQC scheme, due to the approximation considered above, i.e. $C_j(R,t)\simeq C_j(t)$. Even though the delocalisation of the nuclear wave packet is accounted for in a description in terms of multiple independent trajectories, the classical density does not develop a double-peak behaviour but is always centred at the mean nuclear position. This is shown in Fig.~\ref{fig: Figure2}, where the exact nuclear density (black line), calculated from $\Psi(r,R,t)$, is compared to the nuclear density reconstructed from the distribution of classical positions (red line). In the figure, the dashed vertical line indicates the mean nuclear position calculated using $\Psi(r,R,t)$.

According to previous analysis,~\cite{steps,long_steps} the splitting of the nuclear wave packet is caused by the appearance of a step in the TDPES that is strictly related to the spatial dependence of $|C_j(\dulR,t)|^2$: the step producing the splitting appears at the position where $|C_1(\dulR,t)|^2=|C_2(\dulR,t)|^2$. In our MQC approach, this dependence has been neglected and the splitting of the nuclear wave packet is not properly reproduced. Further developments will require an adequate treatment of this spatial dependence in the electronic evolution equation.

\section{Conclusions}
We have shown that the exact factorisation of the electron-nuclear wave function is a promising starting point for the development of approximated MQC schemes to deal with non-adiabatic processes. The approach proposed in this paper is the lowest order approximation to the full quantum mechanical problem. It represents a first attempt toward the development of a MQC method where the approximations can be introduced step-by-step, starting from the exact formulation. In the case presented here, the ``parameter'' $\hbar$ is used to tune the quantum-to-classical approximation: higher order terms can be easily included in our scheme, to go beyond the purely classical approximation of nuclear dynamics. It is interesting to notice that some well-known results can be derived and refined in our formulation, as the role of the classical momentum in inducing electronic non-adiabatic transitions. Furthermore, the exact factorisation provides the exact electronic back-reaction in the form of time dependent scalar and vector potentials, that lead to the derivation of a well defined classical force: it contains (i) a purely ``electromagnetic'' term, representing the direct effect of the electrons on the nuclei, and (ii) an indirect contribution, appearing as an additional inter-nuclear force. Further developments will focus on investigating the properties of this force on a wide range of situations, e.g. when nuclear quantum effects are not negligible, and testing its effect under different conditions, e.g. in the presence of conical intersections.

\section*{Acknowledgements}
Partial support from the Deutsche Forschungsgemeinschaft (SFB 762) and from the European Commission (FP7-NMP-CRONOS) is gratefully acknowledged.

%

\end{document}